# The Electric Discharge in Superhigh Density Gas at Current Amplitude up to 5·10$^5$ A

A.A. Bogomaz, A.V. Budin, M.E. Pinchuk, Ph.G. Rutberg, A.F. Savvateev

*Institute of Problems of Electrophysics of Russian Academy of Sciences (IPE RAS)*
*Dvortsovaya nab. 18, St.-Petersburg, 191186, Russia*
*phone (812)-117-66-23, fax (812)-117-50-56*
*rutberg@iperas.spb.su*

*Abstract*

The investigations of powerful discharge in high density hydrogen were carried out on an installation with preliminary adiabatic compression. The experiments were performed under the following conditions: stored energy of the capacitive storage – 140-450 kJ, charging voltage – 8.0-14.0 kV, discharge current amplitude – 200-600 kA, current rise rate – $10^9$-$10^{10}$ A/s. The maximum particle density $n_{max}$, achieved just before the discharge, was about $2.0 \cdot 10^{22}$ cm$^{-3}$. It was observed one or several compression of the discharge channel. The moments of contractions corresponds a increase of voltage on discharge gap and feature on a curve of a current.

Arc parameters were calculated on the basis of the channel model. It was established that increase of the initial density of particles at increase of a current rate of rise results in reduction of arc diameter with increase of channel temperature.

## Introduction

The results of investigations of powerpulse electric discharge in hydrogen with preliminary adiabatic compression at current amplitude $(1-2) \cdot 10^5$ A are presented in [1, 2].

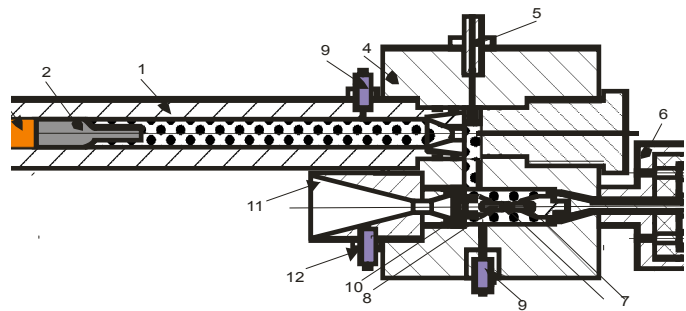

**Fig. 1.** Design of the experimental installation. 1- compression channel, 2 – piston, 3 – powder charge, 4 -discharge chamber, 5 - gas input, 7 - cathode, 8 - grounded anode, 9, 12 - pressure transducers, 10 – diaphragm, 11 - exhausting unit

Modernization of experimental installation resulted on increasing current rate of rise from $10^9$ A/s to $10^{10}$ A/s. Current amplitude was driven to $5 \cdot 10^5$ A. The diagram of a new discharge chamber design is presented on fig. 1. The construction of a diagnostic window 3 mm diameter, made of high impact strength organic, was designed for optical measurements. It was permited to get photostreak of discharge channel and its photo.

## Experimental results

The discharge was initiated after the adiabatic compression up to 100–180 MPa in the discharge chamber just before the discharge. It was corresponded to maximal hydrogen molecules concentration above $10^{22}$ cm$^{-3}$. Oscillogram of pressure in the discharge chamber is showed in fig. 2.

The dynamic volt–ampere character is changed at current amplitude increasing up to $5 \cdot 10^5$ A at increasing its rate of rise. Oscillograms for the electric current and voltage in the discharge gap are presented in fig. 3 for different current rate of rise. It was found that the strength of field and near electrode voltage drop values increase with initial gas density [1, 2] and with increasing of current rate of rise (fig. 4, 5). The values of near electrode voltage drop were obtained from extrapolation interelectrode gap length on zero.

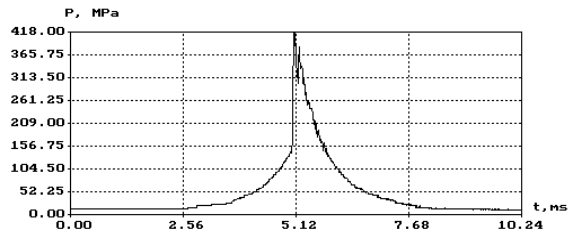

**Fig. 2.** Oscillogram of pressure in the discharge chamber for dJ/dt – $10^{10}$ A/s. The adiabatic compression to pressure of about 150 MPa occur before to time 5.1 ms. After that the electric discharge is initiated.

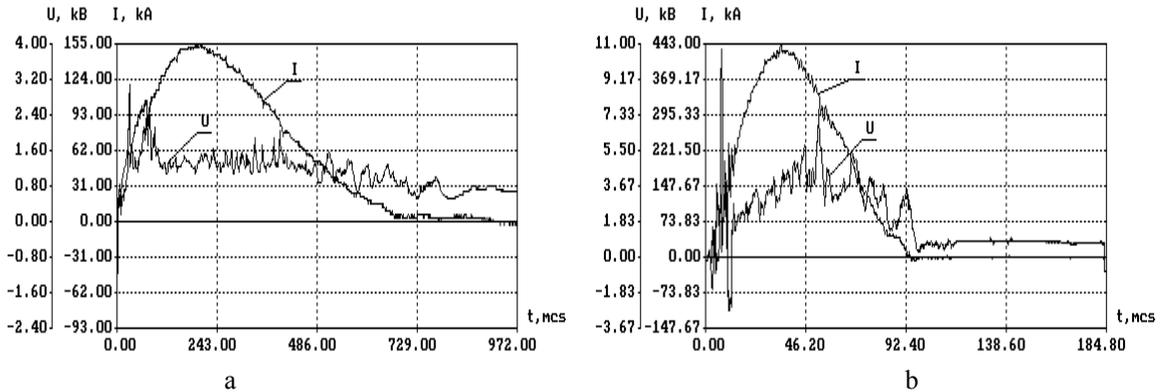

**Fig. 3.** Oscillograms of current and arc voltage. (a – dJ/dt - $10^9$ A/s, b – dJ/dt – $10^{10}$ A/s)

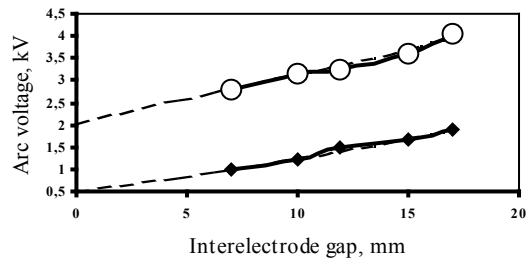

**Fig. 4.** Arc voltage vs interelectrode gap length.. (♦ – dJ/dt - $10^9$ A/s, O – dJ/dt – $10^{10}$ A/s)

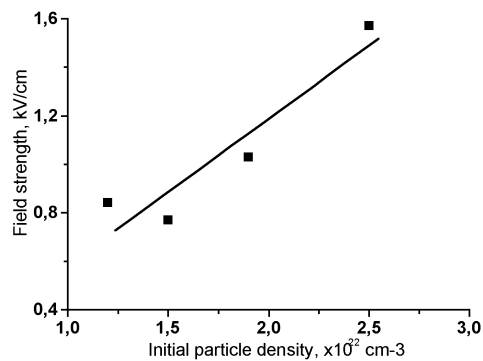

**Fig. 5.** Field strength vs initial particle density.

Photostreak in visible spectrum is presented in fig. 5. Split was perpendicular to the axis in the middle of the discharge gap.

In fig. 6a photostreak of the discharge channel was made at initial hydrogen pressure 5 MPa without preliminary adiabatic compression. Fig. 6b corresponds to the preliminary compression to 110 MPa. In the first case the maximal channel contraction was observed to radius ≤ 0.8 mm. In the second case the minimal radius 1.9 mm, which corresponds to weak glow in the photostreak.

Increasing in arc voltage and dropping (or feature) in the discharge current corresponds to the moment of channel contraction. It ought to be mentioned that sometimes several channel contractions or one but more strong are observed according to the voltage oscillograms (fig. 7, 8). Usually time from the discharge initiation to channel contraction increases at increasing of hydrogen initial particle density before the discharge. Channel contraction can take place both before and after current maximum. At the moment of maximum channel contraction the intensity of radiation in visible part of spectrum decreases sharply

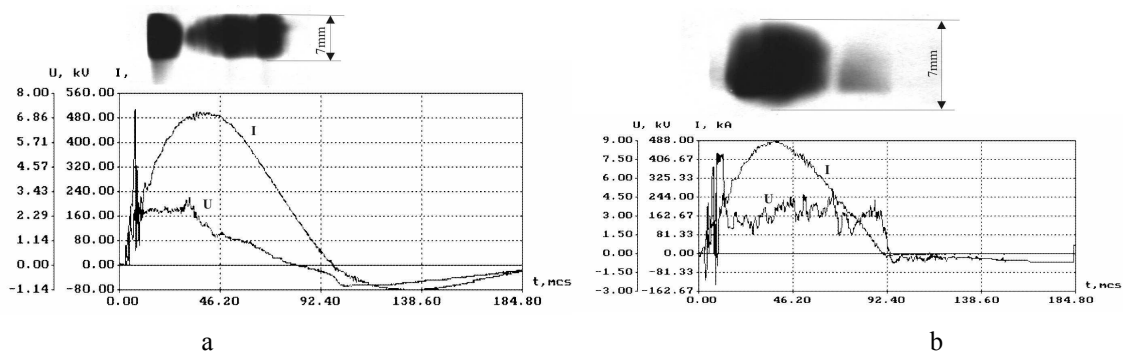

**Fig. 6.** Photostreak, current and voltage: a – initial hydrogen pressure 5 MPa, b – 110 MPa.

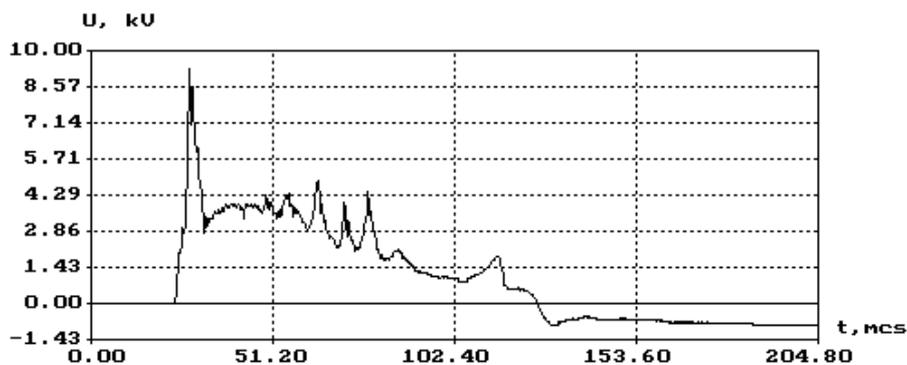

**Fig. 7.** Arc voltage at initial particle density $n_0=1.9\times10^{22}$ cm$^{-3}$ (interelectrode gap is 10 mm).

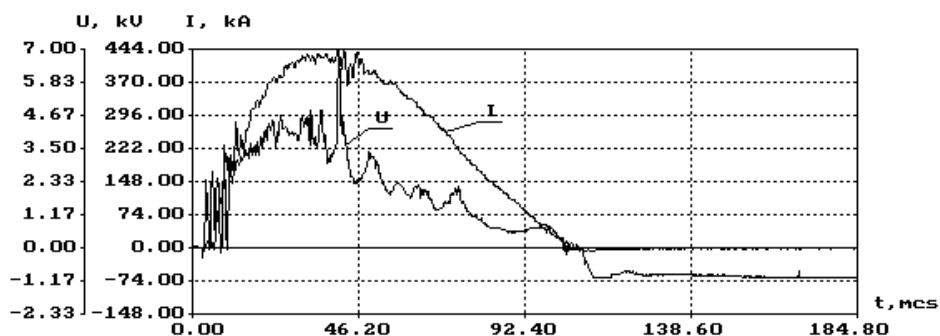

**Fig. 8.** Arc voltage and discharge current at initial particle density $n_0=1.5\times10^{22}$ cm$^{-3}$ (interelectrode gap is 10 mm).

## Discussion

The discharge channel parameters were estimated in two extreme cases: the discharge burns in pure hydrogen (initial concentration of hydrogen is $\sim10^{22}$ cm$^{-3}$) and discharge burn in metal vapour of initiating wire.

In the first case the radiation in the channel is . Plasma is transparent. The system of equations has the following form:

$$\begin{cases} IE = (A n_i^2 T_0^{\frac{1}{2}} + B n_i T_0^{-\frac{1}{2}})\pi r_0^2 \\ 2n_0 kT_0 = P + C\dfrac{I^2}{r_0^2} \\ \dfrac{E}{I} = \dfrac{1}{\pi r_0^2 \sigma(T_0, n_0)} \\ n_i = n_e \sim n_0 \end{cases}$$

here T – channel temperature, $n_i$ – ion concentration, $r_0$ – channel radius, σ – conductivity, P – pressure after adiabatic compression, I – current, E – strength of field.

Values *P, I* and *E*, systems necessary for the system decision, are took from oscillograms at the moment of time corresponding to the maximal contraction. Calculated values of conductivity for hydrogen and metal plasma took from works [3] and [4] accordingly. The system of the equations was solved for two values of strength of a field. At more frequently repeating value *E* = 1600 V/sm to the appropriate average contractions (fig. 5) and at maximal contraction *E* = 3600 V/cm. Thus in both cases *J* = 400 kA and *P* = 300 MPa. The system decision for the first case
*E* = 1600 V/cm: *T* = 3·10$^5$ K, *n0* = 3.4·10$^{20}$ cm$^{-3}$, *r* = 0.1 cm
*E* = 3600 V/cm: *T* = 3·10$^5$ K, *n0* = 1.0·10$^{21}$ cm$^{-3}$, *r* = 0.06 cm
Accuracy of the system decision is ~10 %.

It is visible, that to deeper contraction there corresponds the higher density and smaller radius of the channel. Thus the total number of particles in the channel is remained.

In that case if the discharge burns in metal vapour, then the optical density of the channel is great. Therefore it was considered, that it radiates as black body.

In the second case the optical density of the channel is high. Therefore it was considered that it radiates as a black body.

In this case from the system:

$$\begin{cases} IE = 2\pi r_0 \sigma_b T^4 \\ (1+\bar{z})kT_0 = P + C\dfrac{I^2}{r_0^2} \\ \dfrac{E}{I} = \dfrac{1}{\pi r_0^2 \sigma(T_0, n_0)} \end{cases}$$

We receive result $r_0$ = 0.2 cm, *T* = 1.25·10$^5$ K, $n_0$ = 10$^{20}$ cm$^{-3}$.

**Conclusion**

Estimations show that experiment corresponding conditions in fig. 5a is more suitable to burning arc in hydrogen. The condition in fig. 5b closely corresponds to arc burning in initiating wire vapour, its radius of expansion on this condition is 0.2 cm according to independent estimations and close to measured channel radius.

Field strength and near electrode voltage drops are increased with current rise rate and initial particle density increase.

This work is partially supported by the Russian Fund of Fundamental Investigation (RFFI), project N 02-02-16770.